\documentclass[preprint,12pt]{elsarticle}
\usepackage{amssymb}
\usepackage{bm}
\usepackage{mathtools}
\usepackage{subcaption}

\makeatletter
\newcommand{\currentfsize}{\f@size pt}
\makeatother

\usepackage{graphicx}
\usepackage{array}
\usepackage{amsmath}
\usepackage[nodisplayskipstretch]{setspace}
\usepackage{wrapfig,booktabs}

\journal{Journal of Sound and Vibration}

\begin{document}
\begin{frontmatter}
\title{Vold-Kalman Filter Order tracking of Axle Box Accelerations for Railway Stiffness Assessment}
%
%
\author[label1]{Cyprien Amadis Hoelzl\corref{cor1}}
\ead{hoelzl@ibk.baug.ethz.ch}

\author[label1]{Vasilis Dertimanis}
\author[label2]{Lucian Ancu}
\author[label2]{Aurelia Kollros}
\author[label1]{Eleni Chatzi}
            
\address[label1]{Department of Civil, Environmental and Geomatic Engineering, ETH Z{\"u}rich}
\address[label2]{Measurement and Diagnostics, SBB, Bern, Switzerland}

%

\begin{abstract}
Intelligent data-driven monitoring procedures hold enormous potential for ensuring safe operation and optimal management of the railway infrastructure in the face of increasing demands on cost and efficiency. 
Numerous studies have shown that the track stiffness is one of the main parameters influencing the evolution of degradation that drives maintenance processes.
As such, the measurement of track stiffness is fundamental for characterizing the performance of the track in terms of deterioration rate and noise emission. This can be achieved via low-cost On Board Monitoring (OBM) sensing systems (i.e., axle-box accelerometers) that are mounted on in-service trains and enable frequent, real-time monitoring of the railway infrastructure network. Acceleration-based stiffness indicators have seldom been considered in monitoring applications.
In this work, the use of a Vold-Kalman filter is proposed, for decomposing the signal into periodic wheel and track related excitation--response pairs functions. We demonstrate that these components are in turn correlated to operational conditions, such as wheel out-of-roundness and rail type. 
We further illustrate the relationship between the track stiffness, the measured wheel-rail forces and the sleeper passage amplitude, which can ultimately serve as an indicator for predictive track maintenance and prediction of track durability.

\end{abstract}


\end{frontmatter}

%
\section{Introduction}

Track stiffness is a primary parameter for characterizing the performance of railway infrastructure in terms of condition degradation \cite{Nielsen2020} and noise emission to the surrounding environment \cite{oegren2006}. The measurement of stiffness is considered a fundamental driver for continued development of railway engineering, holding both theoretical and practical significance \cite{Wang2016}.
The degradation of railways tracks is typically modelled on the basis of an analytical degradation model (e.g., a settlement equation) that relates cycles of dynamic forcing to incremental damage accumulation (i.e., ballast settlement or component wear) \cite{Dahlberg2001}; track stiffness is one of the main parameters influencing loads in such degradation models. Statistical degradation models are often preferred to the aforementioned mechanistic approach, because statistical models can better capture the large uncertainty in the track geometry behavior \cite{Soleimanmeigouni2018}.

In this context, Real et al. \cite{Real2012} compare the analytic solution of a Zimmerman beam on an elastic bedding to a Finite Element (FE) based analysis of a track section, to 
demonstrate that higher track stiffness, or equivalently low subsidence, results in axle loads that are distributed between fewer sleepers. As a result, the rail--wheel and sleeper-ballast contact forces increase, which can lead to premature rail and ballast wear \cite{Tzanakakis2013}. A sudden decrease in track stiffness is often related to local damage to the ballast, substructure, and subgrade (changes in track type, fouled ballast, mud pumping or hanging ties) \cite{Hoelzl2020}.
Hence, within a predictive maintenance scheme, the assessment of track stiffness allows for suited maintenance actions, such as grouting or placement of under-sleeper pads for reducing wheel/rail contact forces, which aim to the durability of the system \cite{Dahlberg2010}.

Track stiffness can be measured by standstill measurements at single track locations or by vehicle--based systems. Standstill measurements are obtained by measuring the rail deflection due to the axle load of passing trains using sensors on the track \cite{Wang2016}. 
Continuous stiffness measurements from specialized vehicles are often preferred over standstill measurements, because they deliver an assessment along extended track lengths.
The collection of data from diagnostic and in-service vehicles enables the network wide assessment of the current (diagnosis) and future (prognosis) state of the assets. Diagnostic vehicles regularly measure the condition of the track with high accuracy, albeit due to their complexity of operation with limited temporal resolution. Diagnostic vehicle--based stiffness measurements often use the subsidence of the track under an axle load as an indicator of stiffness \cite{Soldati2006}. On-Board Monitoring (OBM) vehicles have recently been introduced in order to support the sparser measurements from diagnostic vehicles \cite{HoelzlReview2021}. OBM vehicles are passenger vehicles equipped with sensors such as Axle Box Accelerometers (ABA) that enable a nearly daily data collection. Such acceleration data must first be processed into quantifiable track quality indicators \cite{Yan2020}. Several methods have in the past been successfully applied in order to relate accelerations to geometric flaws by using Kalman filters \cite{Dertimanis2020} or double integration and filtering techniques \cite{HoelzlReview2021}. Non-parametric methods such as time frequency analysis using the Discrete Wavelet Transform \cite{hoelzlIMACXL} or the Continuous Wavelet Transform \cite{Molodova2015} have been applied successfully applied to identify rail flaws such as squats. 
The track stiffness and wheel out-of-Roundness result in non-stationary response. 
Linear Parameter Varying AutoRegressive models (LPV-AR) can capture the non-stationary vibration response for varying superstructure stiffnesses when the dynamics are controlled by an external scheduling variable such as the vehicle speed \cite{HoelzlIMAC2020}. The amplitude of the wheel out-of-roundness has been in the past characterized using the Empirical Mode Decomposition (EMD) and the Hilbert-Huang transform (HHT) \cite{Song2020}.
Quirke et al. propose a method to estimate the track stiffness using the vertical bogie acceleration in combination with a vehicle track interaction model and cross--entropy optimization \cite{Quirke2017}. Their methodology has been verified on simulated data, however, the practical usability of the methodology may be limited due to the simplified nature of the model and the uncertainty of the physical parameters.

In suggesting a scheme that is fit for an on--board, automated setting, we propose a stiffness indicator based on use of a Vold-Kalman Filter (VKF) on Axle Box Acceleration (ABA) measurements. Unlike standstill stiffness and subsidence measurements, the proposed indicator can be applied in an in-service monitoring scheme, which attempts to deliver continuous assessment. 
The VKF is a parametric identification method that can be used to decompose the signal into harmonic responses, or orders, of periodic loads \cite{Herlufsen199934}\cite{specsheetBruelKjaer}. The VKF has been extensively used for monitoring the condition of rotating machinery equipment such as wind turbine gearboxes \cite{Hong2018}\cite{Li2019}. 

In the current setting, the VKF is used to extract the response functions corresponding to the sleeper passage frequency and to the harmonic wheel out of roundness frequencies from ABA.
We demonstrate that these response functions are related to the operational conditions, such as subsidence (e.g. infrastructure type), wheel out of roundness and vehicle speed. We further show the relation between track stiffness, rail--wheel forces and the sleeper passage amplitude, suggesting that a VKF-derived stiffness indicator can indeed support maintenance actions in a straightforward manner.

The structure of this paper is organized as follows: In Sec.~\ref{sec:MRTS} we offer a literature review of track stiffness monitoring approaches using direct stiffness measurements or indirect assessments via acceleration measurements. The fundamental aspects of the Vold-Kalman Filter are described in Sec.~\ref{sec:VKF}, whereas Sec.~\ref{sec:CSonABA} contains the measurement data description and the results of the application of the proposed stiffness and wheel Out-Of-Roundness identification. Finally, Sec.~\ref{sec:Concl} outlines the core concept and results of the proposed methodology.
\section{Monitoring Railway Track Stiffness}
\label{sec:MRTS}

\subsection{Direct Stiffness Measurements} 

As aforementioned, vehicle--based stiffness measurements are generally preferred to standstill ones. Diverse approaches exist for static and dynamic track stiffness assessment. The static stiffness of the track structure results from the deformation $U_0$ of the track under a static load $Q_0$, while the dynamic stiffness corresponds to the stiffness under a dynamic excitation.  The subsidence measurement vehicle, employed by the Swiss Federal Railways (EMW from InfraMT), uses a heavy car (20 tons) connected to a light car of negligible mass. The static stiffness of the track is estimated by measuring the deflection (subsidence) of the track under crossings of the heavy car from the reference point of a light car \cite{Soldati2006}. The track stiffness measurement method developed at the University of Nebraska-Lincoln (USA) \cite{Norman2004} adopts a similar principle, involving a laser distometer mounted at an offset to the vehicle axle to measure the deflection of the bogie with respect to the unloaded track. These two vehicles require a relatively slow speed during the measurement (below 40~km/h).
The CETE (Centre d'Experimention et de Recherche, France) and the Engineering Department of SNCF (Paris) has developed a stiffness measurement vehicle (D2.1.9 Innotrack 2009) using MEMS sensors \cite{Hosseingholian2006}. The wheels on the measurement axle are mounted off-centered on the axle and this imbalance results in a sinusoidal excitation of the system. Vertical accelerometers mounted the bogie and the axle are then exploited to compute the force exerted by the system on the rail. The vertical displacement of the system is obtained via double integration of the axle acceleration. The dynamic stiffness of the track is calculated from the resulting force-displacement hysteresis curve. Tab.~\ref{tab:00} summarizes such established measurement concepts for measuring track stiffness, along with the herein proposed concept of indirectly assessing stiffness through acceleration measurements, which is elaborated in the next section.

\begin{table}[h]
\caption{Summary of track stiffness measurement schemes}\label{tab:00}%
\centering
\small
\begin{tabular}{p{3.2cm}p{3.0cm}p{3.0cm}p{1.4cm}p{0.7cm}}
\hline
Concept & Advantages & Disadvantages & Speed & Refs. \\ \hline
Standstill measurements & Mature technology  & Limited scalability & Slow speed & \cite{Wang2016}  \\[0.8cm]
Continuous measurements of static stiffness under axle load&          Mature technology  & Accuracy may be lower than standstill & Under 40~km/h      &  \cite{Soldati2006} \cite{Wangqing1997}         \\[0.8cm]
Continuous measurement of dynamic stiffness via eccentric wheel excitation  & Identification of both stiffness and damping, hysteresis curve & Low speed, research stage              & 6~km/h  & \cite{Hosseingholian2006} \\[0.8cm]
Acceleration based assessment of dynamic stiffness            & Identification of both stiffness and damping          & Experimental systems  & 200~km/h      &  \cite{Quirke2017}          \\ \hline
\end{tabular}
\end{table}
\raggedbottom

\subsection{Indirect assessment of dynamic stiffness via ABAs}
The use of specialized diagnostic vehicles as a means for OBM delivers high quality measurements, but restricts the possibility of delivering truly continuous network diagnostics. Such vehicles feature highly specialized and thus costly equipment, while at the same time their runs are restricted to few times a year, since they come with costs and restrictions in the availability of the railway network.
On the other hand, it is possible to employ in-service vehicles as a means to OBM. In this case, simple lower cost sensors can be exploited (e.g., only axle box accelerometers), and while the quality of data aggregated from a single run is decreased, the benefit lies in the utilization of multiple vehicles for OBM purposes, thus achieving a spatially and temporally dense supervision of the network. The collected data from such in-service OBM vehicles can serve for assessing both the track and vehicle condition.

In this scenario, the primary source of information comes in the form of acceleration time series. Condition indicators are often extracted from time series by using parametric or non-parametric methods. Parametric methods rely on system parameters to describe the system dynamics \cite{HoelzlReview2021}.
Dertimanis et al. \cite{Dertimanis2020} use a half-car model to accurately reconstruct the longitudinal level from axle box accelerations with a kalman filter. The track stiffness estimation methodology proposed by Quirke et al., combines a vehicle track interaction model with a cross--entropy optimization technique \cite{Quirke2017}.

\begin{figure}[!ht]
  \begin{center}
  \includegraphics[trim={0 0.3cm 0 0.3cm},clip]{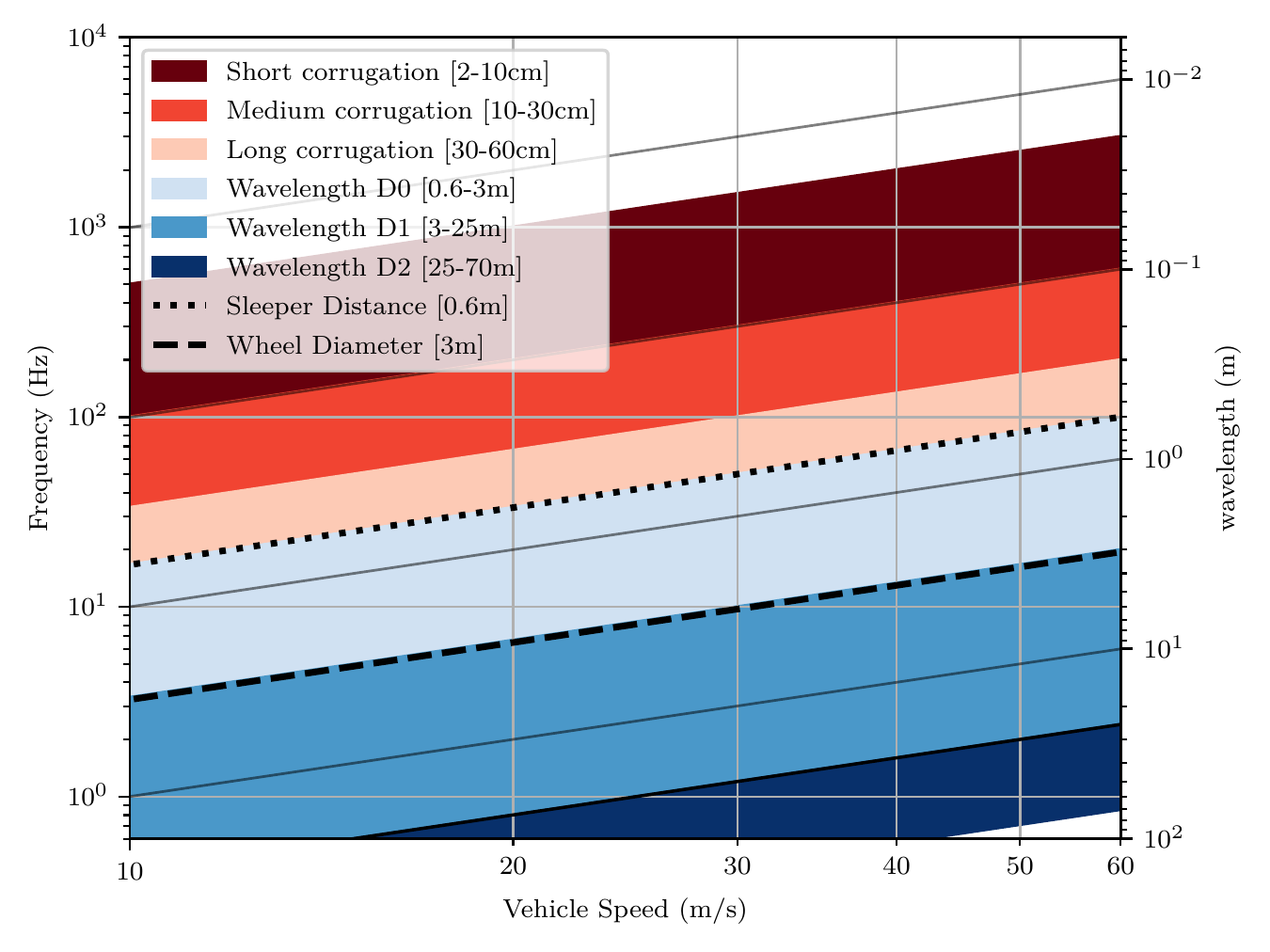}
  \end{center}
\caption{Relation between vehicle speed, frequency and wavelength ~\cite{HoelzlReview2021}}
\label{fig:000}
\end{figure}

The previous approaches rely on availability and use of a physical model, even if approximate. This particular problem also admits a purely data-driven approach for condition assessment, as dictated by the nature of applied excitation. The wheelset is excited by geometric irregularities of the track \cite{Steenbergen2016}, out-of-roundness of the wheel \cite{Steenbergen2013} and parametric excitation (sleeper passage frequency, track parameters). Each of these excitation sources comprises a specific wavelength range (e.g. sleeper spacing and factors of the wheel diameter). The vehicle speed is, as illustrated in Fig.~\ref{fig:000} \cite{HoelzlReview2021}, the main factor modulating the excitation frequencies relating to these effects.
Several signal processing methods have been proposed to identify periodic excitations from non-stationary signals. Song et al. \cite{Song2020} characterize the  wheel out-of-roundness by applying the Empirical Mode Decomposition and the Wigner-Ville distribution to ABA. The Hilbert Huang transform spectrum characteristics have been used to study the effect of wheel flats on ABA while proposing different the frequency bands at different running speeds \cite{Li2012}. Li et al. propose a short pitch corrugation detection method based on signatures in the wavelet power spectrum \cite{Li2015}. The vehicle speed has a significant impact on the frequency bands of interest, but the most commonly used signal processing methods reviewed previously are not directly conditioned by this essential parameter.
The next section describes the methodology that we propose for decomposing the collected signals into periodic versus random excitation components 
while taking into account the non-stationary vehicle velocity.

\section{The Vold-Kalman Filter}
\label{sec:VKF}
\subsection{The process and measurement equations}
Given a realization of a discrete-time, noise-corrupted ABA stochastic process, the Wold decomposition reads \cite{Wold1938},
\begin{equation}\label{eq:1}\
y[k] = x[k] + \eta[k]
\end{equation}
where $k\in\mathbb{Z}^+$, $t=kT_s$~(s) are the associated discrete time indices for a given equidistant sampling period $T_s$, and $x[k]$, $\eta[k]$ comprise the mutually uncorrelated deterministic and stochastic components, respectively, with the latter corresponding to broadband random noise. 

For the aims of the current study, $x[k]$ is assumed to form a superposition of harmonic components, which are represented as modulated carrier waves of varying frequency~\cite{Tuma2005}, parameterized over the vehicle speed $v(t)$. That is
\begin{equation}\label{eq:2}
x[k]=  \sum_{n=1}^sA_{n}[k]e^{j\phi_n[k]}
\end{equation}
where $A_{n}[k]$ is a complex envelope and   
\begin{equation}\label{eq:3}
\phi_n[k]=\sum_{\ell=1}^{k} 2\pi f_n[\ell]T_s
\end{equation}
is a complex phasor, with $f_n[\ell]$ denoting the instantaneous frequency in Hz.

The Vold-Kalman filter aims at identifying the complex envelopes $A_n[k]$, given noise-corrupted observations $y[k]$, for $k=1,\dots,N$, and estimates of the instantaneous frequencies $f_n[\ell]$. Assuming that the complex envelopes comprise \textit{smooth}, \textit{slowly-varying} modulations of their associated phasors, they may be approximated by   
\begin{equation}\label{eq:4}
\nabla^qA_n[k]=\epsilon_{n,q}[k]
\end{equation}
where $\nabla^q$ is the finite difference operator of order $q$ and $\epsilon_{n,q}[k]$ may be considered as the $q$-th gradient of $A_n[k]$. Under the adopted assumptions, Eq.~\ref{eq:4} implies that the finite differences among successive time instants of the complex envelopes are ``small''. Moreover, it introduces a filtering effect. Indicatively, for $q=1,2,3$ we get,
\begin{subequations}
	\begin{align}
		A_n[k]-A_n[k-1] &= \epsilon_{n,1}[k] \label{eq:5a}\\
		A_n[k]-2A_n[k-1]+A_n[k-2] &= \epsilon_{n,2}[k] \label{eq:5b}\\
		A_n[k]-3A_n[k-1]+3A_n[k-2]-A_n[k-3] &= \epsilon_{n,3}[k] \label{eq:5c}
	\end{align}
	\label{eq:5}%
\end{subequations}
which pertain to standard, non-homogeneous difference equations, represented by associated fixed-pole digital transfer functions in the $\mathcal{Z}$ domain.

By considering Eqs.~\ref{eq:4} and~\ref{eq:1} as \textit{process} and \textit{measurement} equations, respectively, a state-space system of the form,
\begin{subequations}
	\begin{align}
		\nabla^q\mathbf{a}[k] &= \bm{\epsilon}[k] \label{eq:6a}\\
		y[k] &= \mathbf{c}[k]\mathbf{a}[k]+\eta[k] \label{eq:6b}
	\end{align}
	\label{eq:6}%
\end{subequations}
is generated, where 
\begin{equation}\label{eq:7}
    \mathbf{a}[k]=\Big[A_1[k]~A_2[k]~\dots~A_s[k]\Big]^T\qquad [s\times1]
\end{equation}
is the unknown state vector,
\begin{align}
    \bm{\epsilon}[k] &= \Big[\epsilon_{1,q}[k]~\epsilon_{2,q}[k]~\dots~\epsilon_{s,q}[k]\Big]^T\qquad [s\times1]\label{eq:8}\\
    \mathbf{c}[k] &= \Big[e^{j\phi_1[k]}~e^{j\phi_2[k]}~\dots~e^{j\phi_s[k]}\Big]\qquad [1\times s]\label{eq:9}
\end{align}
and $\phi_n[k]$ is given by Eq.~\ref{eq:3}.
\subsection{A least-squares solution}
We cannot apply the original Kalman filter to the set of Eqs.~\ref{eq:6}, since both $\bm{\epsilon}[k]$ and $\eta[k]$ are unknown and the entries of $\mathbf{c}[k]$ are usually estimated from data. Instead, a Kalman \textit{smoothing} solution can be established on the basis of available measurements~\cite{Feldbauer2000}. To demonstrate this solution, define the $[N\times 1]$ vectors
\begin{equation}\label{eq:10}
\mathbf{A}_n=
    \begin{bmatrix}
    A_n[1]\\
    A_n[2]\\
    \vdots\\
    A_n[N]
    \end{bmatrix},\qquad
\mathbf{E}_n=
    \begin{bmatrix}
    \epsilon_{n,q}[1]\\
    \epsilon_{n,q}[2]\\
    \vdots\\
    \epsilon_{n,q}[N]
    \end{bmatrix}
\end{equation}
for $n=1,2,\dots,s$ and
\begin{equation}\label{eq:11} 
\mathbf{y}=
    \begin{bmatrix}
    y[1]\\
    y[2]\\
    \vdots\\
    y[N]
    \end{bmatrix},\qquad
\bm{\eta}=
    \begin{bmatrix}
    \eta[1]\\
    \eta[2]\\
    \vdots\\
    \eta[N]
    \end{bmatrix}
\end{equation}
Then, Eq.~\ref{eq:4} can be written over the available data indices as
\begin{equation}\label{eq:12}
\mathbf{S}_q\mathbf{A}_n=\mathbf{E}_n
\end{equation}
for $n=1,2,\dots,s$, where $\mathbf{S}_q$ is a \textit{known} $[N\times N]$ matrix that depends only on the filter's order $q$ (i.e., it is $n$-independent, as demonstrated in Eq.~\ref{eq:5}). By further defining the $[Ns\times 1]$ vectors
\begin{equation}\label{eq:13}
\mathbf{a}=
    \begin{bmatrix}
    \mathbf{A}_1\\
    \mathbf{A}_2\\
    \vdots\\
    \mathbf{A}_s
    \end{bmatrix},\qquad
\mathbf{e}=
    \begin{bmatrix}
    \mathbf{E}_1\\
    \mathbf{E}_2\\
    \vdots\\
    \mathbf{E}_s
    \end{bmatrix}
\end{equation}
allows expanding Eq.~\ref{eq:12} over the index $n$ as
\begin{equation}\label{eq:14}
\mathbf{S}\mathbf{a}=\mathbf{e}
\end{equation}
for $\mathbf{S}=\mathbf{I}_s\otimes\mathbf{S}_q$.

Accordingly, Eq.~\ref{eq:6b} may be written as
\begin{equation}\label{eq:15}
    y[k]=\sum_{n=1}^sc_n[k]A_n[k]+\eta[k]
\end{equation}
with $c_n[k]$ denoting the $n$-th entry of $\mathbf{c}[k]$ in Eq.~\ref{eq:9}. Writing Eq.~\ref{eq:15} for $k=1,2,\dots,N$ gives
\begin{align*}
    y[1]   &= c_1[1]A_1[1] +c_2[1]A_2[1]+\dots+c_s[1]A_s[1]+\eta[1]\\
    y[2]   &= c_1[2]A_1[2] +c_2[2]A_2[2]+\dots+c_s[2]A_s[2]+\eta[2]\\
    \vdots & \qquad\qquad\qquad\qquad\qquad\vdots\\
    y[N]   &= c_1[N]A_1[N] +c_2[N]A_2[N]+\dots+c_s[N]A_s[N]+\eta[N]
\end{align*}
or, using the definitions of Eqs.~\ref{eq:10}--\ref{eq:11}
\begin{equation}\label{eq:16}
    \mathbf{y}=\sum_{n=1}^s\mathbf{C}_n\mathbf{A}_n+\bm{\eta}
\end{equation}
with $\mathbf{C}_n=\text{diag}\{c_n[1],c_n[2],\dots,c_n[N]\}$. Then, for
\begin{equation}\label{eq:17}
\mathbf{C}=
    \begin{bmatrix}
    \mathbf{C}_1 & \mathbf{C}_2 & \dots & \mathbf{C}_s
    \end{bmatrix}
\end{equation}
Eq.~\ref{eq:16} becomes
\begin{equation}\label{eq:18}
\mathbf{y}=\mathbf{C}\mathbf{a}+\bm{\eta}
\end{equation}

A least-squares problem can be now formulated by considering the objective function
\begin{equation}\label{eq:19}
    V(\mathbf{a})=\big(\mathbf{R}\mathbf{e}\big)^T\big(\mathbf{R}\mathbf{e}\big)+\bm{\eta}^T\bm{\eta}
\end{equation}
where $\mathbf{R}=\text{diag}\{r_1,r_2,\dots,r_s\}$ is a diagonal matrix of weighting factors. Substituting Eqs.~\ref{eq:14},\ref{eq:18} and setting the gradient of $V(\mathbf{a})$ equal to zero finally yields
\begin{equation}\label{eq:20}
    \Bigg[\mathbf{C}^H\mathbf{C}+\big(\mathbf{R}\mathbf{S}\big)^T\big(\mathbf{R}\mathbf{S}\big)\bigg]\mathbf{a}=\mathbf{C}^H\mathbf{y}
\end{equation}
with $H$ denoting Hermitian transpose. Due to the special structure of the involved matrices, Eq.~\ref{eq:20} constitutes an ill-conditioned problem, the solution of which requires iterative methods, such as the preconditioned conjugate gradient one. The reader is referred to Feldbauer and Holdrich \cite{FeldBauerHoeldrich2000} for further details.

\subsection{Implementation and solution}
The resulting linear system of coupled equations is formulated as a sparse matrix product and is solved using a sparse direct solver \cite{Davis2004}. The execution time of this solver depends linearly on the matrix order. For a large time series, the direct solution is obtained by dividing the time history into overlapping bins. The decomposed signals are reassembled with a Hanning window taper on the overlapping parts.
The source code for the Python 3 implementation of the second order VKF written as a part of this paper has been made openly available \cite{GithubPyVKF} for reuse by interested readers.

\subsection{Validation on a simulated signal}
\begin{figure}
\centering
  \begin{subfigure}{\textwidth}
    \includegraphics[trim={0.0cm 0.2cm 0 0.2cm},clip]{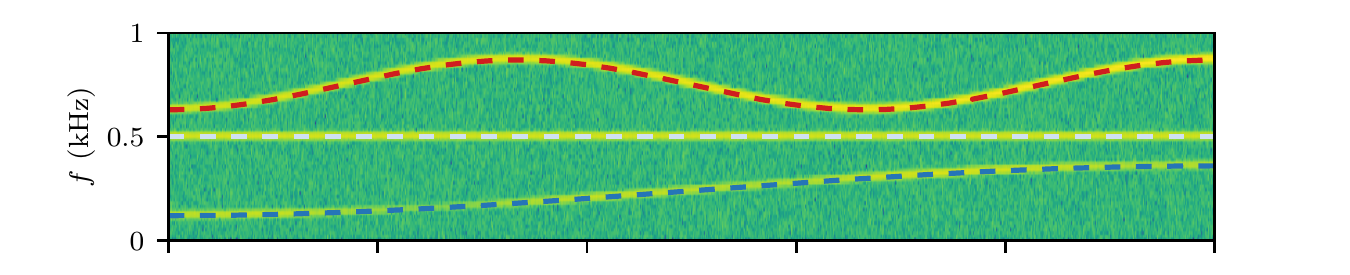}
    \caption{The spectrogram of the original noisy signal $y$ and the frequency of the extracted components.} \label{fig:0a}
  \end{subfigure}
  \begin{subfigure}{\textwidth}
    \includegraphics[trim={0.0cm 0.2cm 0 0.0cm},clip]{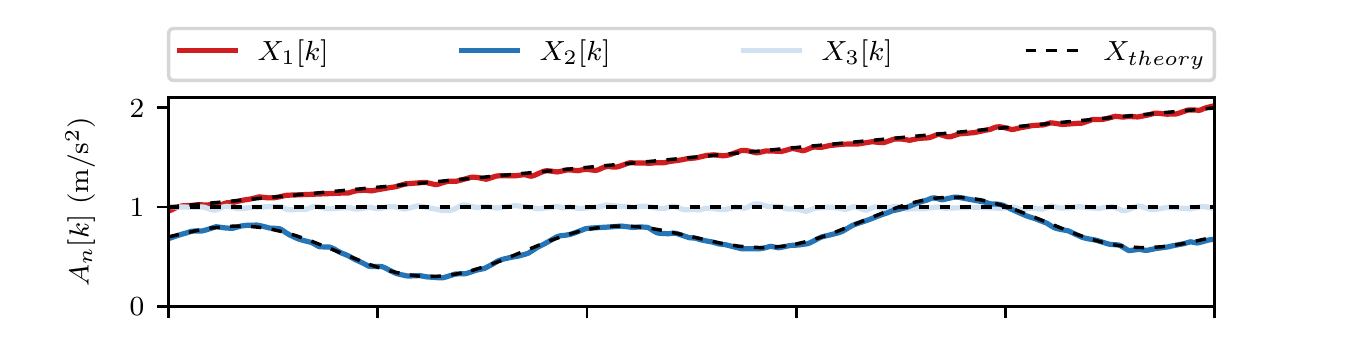}
    \caption{} \label{fig:0b}
  \end{subfigure}
  \begin{subfigure}{\textwidth}
    \includegraphics[trim={0.0cm 0.1cm 0 0.2cm},clip]{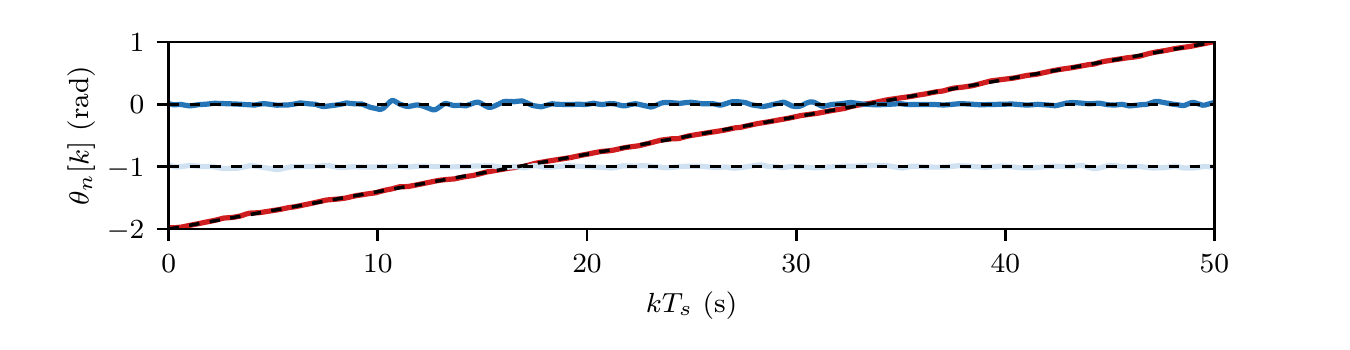}
    \caption{} \label{fig:0c}
  \end{subfigure}

\caption{Wold decomposition of synthetic noisy signals for VKF of order $p=2$. 
Comparison of the extracted parameters to the synthetic signal for (a) xx (b) the amplitude, and (c) the phase.} \label{fig:0}
\end{figure}

To more clearly demonstrate what is described as part of the previous theoretical section, the Vold-Kalman filter is first evaluated on a synthetic signal. Three instantaneous frequencies $f_n$ of order $l=1$ are selected. The observed signal $y[k]$ in Eq.~\ref{eq:21} is the sum of the non-stationary deterministic sinusoidal signals $X_1$, $X_2$ and $X_3$, and Gaussian noise of standard deviation $\sigma_{\eta}=0.75$~m/s$^2$. 
\begin{equation}\label{eq:21}
    y[k]=x + \eta = X_1[k]+X_2[k]+X_3[k] + \eta[k]
\end{equation}
The non-stationary signals are defined with time--varying phase and amplitude as
\begin{subequations}
	\begin{align}
        X_1 &= a_1\cdot cos(2\pi\cdot f_1 t+ p_1) \label{eq:22a}\\
		X_2 &= a_2\cdot cos(2\pi\cdot f_2 t+p_2) \label{eq:22b}\\
		X_3 &= a_3\cdot cos(2\pi\cdot f_3 t + p_3) \label{eq:22c}
	\end{align}
	\label{eq:22}%
\end{subequations}
where the discretized sampling time steps are $t = kT_s$. The signal is sampled at a rate of $f_s = 1/T_s = 12~kHz$. The time varying amplitude $a_n$, frequency $f_n$ and phase $p_n$ of the non-stationary signals are defined as:
\begin{subequations}
	\begin{align}
        \begin{bmatrix}
        a_1\\
        a_2\\
        a_3\\
        \end{bmatrix}&=
        \begin{bmatrix}
        (0.02t+1)\\
        1+0.5\cdot sin(2\pi \cdot0.02t)\cdot cos(2\pi \cdot0.04 t)\\
        1\\
        \end{bmatrix} \label{eq:23a}\\
        \begin{bmatrix}
        f_1\\
        f_2\\
        f_3\\
        \end{bmatrix}&=
        \begin{bmatrix}
        (600+120 \cdot cos(2\pi \cdot0.03t) )\\
        (240-120\cdot cos(2\pi \cdot0.01t)\\
        500\\
        \end{bmatrix} \label{eq:23b}\\
        \begin{bmatrix}
        p_1\\
        p_2\\
        p_3\\
        \end{bmatrix}&=
        \begin{bmatrix}
        0.06t-2\\
        0\\
        -1\\
        \end{bmatrix} \label{eq:23c}
	\end{align}
	\label{eq:23}%
\end{subequations}
One can observe from Eq.~\ref{eq:22} and Eq.~\ref{eq:23} that the first component $X_1$ features a sinusoidal frequency variation and a linearly varying amplitude variation and phase. $X_2$ has a sinusoidal frequency and amplitude variation. $X_3$ corresponds to a stationary signal with constant frequency, amplitude and phase. Figure~\ref{fig:0} shows the characteristics of the frequencies, amplitudes and phases of the signals $X_n$. The solution of the system of equation as defined in Eq.~\ref{eq:18} is enables the separation of the harmonic signals $X_n$ from the remaining white noise $\eta$. Figure~\ref{fig:0b} and Fig.~\ref{fig:0c} show that the amplitudes and phases estimated with the VKF closely match the theoretical ones. The estimated amplitudes and phases enable the reconstruction of the signal.

\begin{figure}
\centering
  \begin{subfigure}{\textwidth}
    \includegraphics[trim={0.0cm 0.2cm 0 0.4cm},clip]{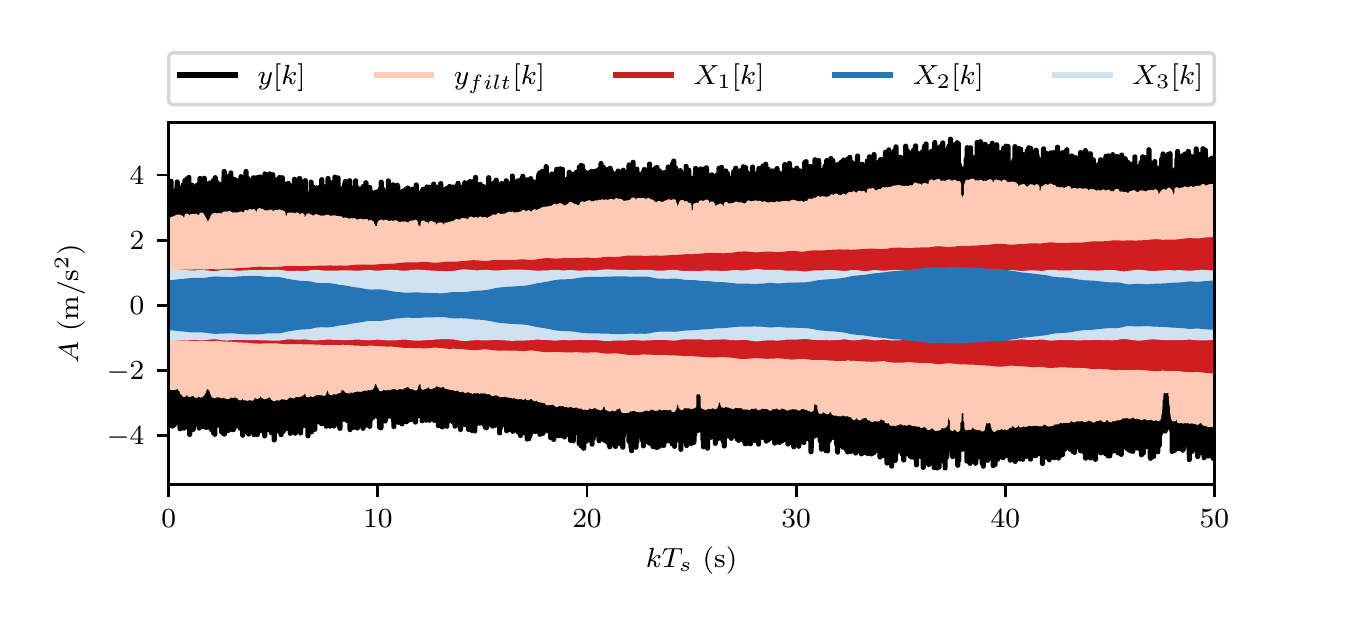}
    \caption{Full time series of the unfiltered signal $y$ and the filtered signals $y_{filt}$, $X_1$, $X_2$ and $X_3$.} \label{fig:1a}
  \end{subfigure}
  \begin{subfigure}{\textwidth}
    \includegraphics[trim={0.0cm 0.2cm 0 0.1cm},clip]{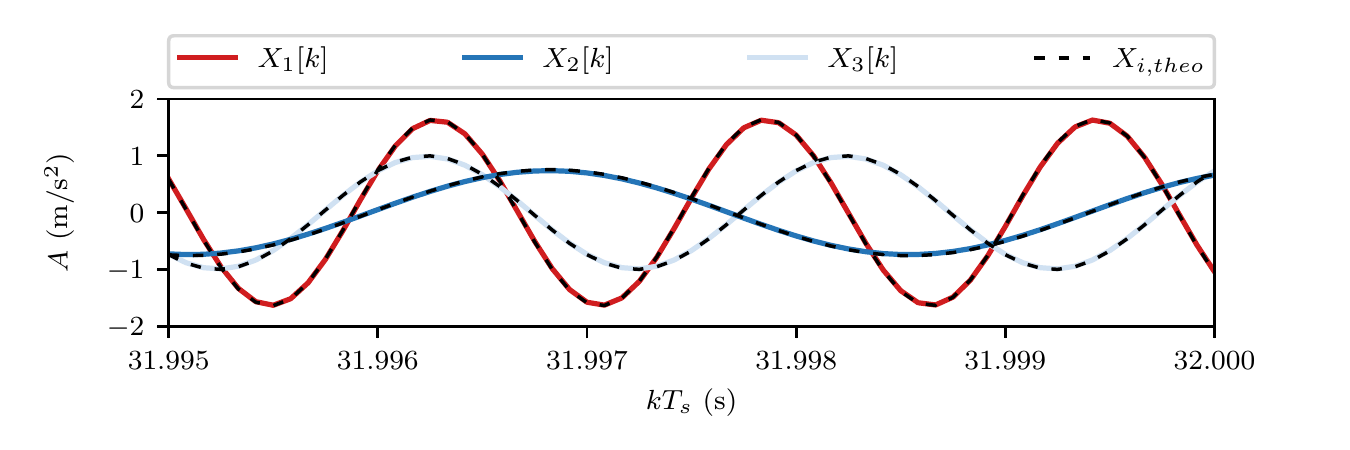}
    \caption{Zoom in view illustrating the very good agreement between the extracted components $X_n$ and the theoretical components $X_{i,theo}$.} \label{fig:1b}
  \end{subfigure}
  \begin{subfigure}{\textwidth}
    \includegraphics[trim={0.0cm 0.2cm 0 0.1cm},clip]{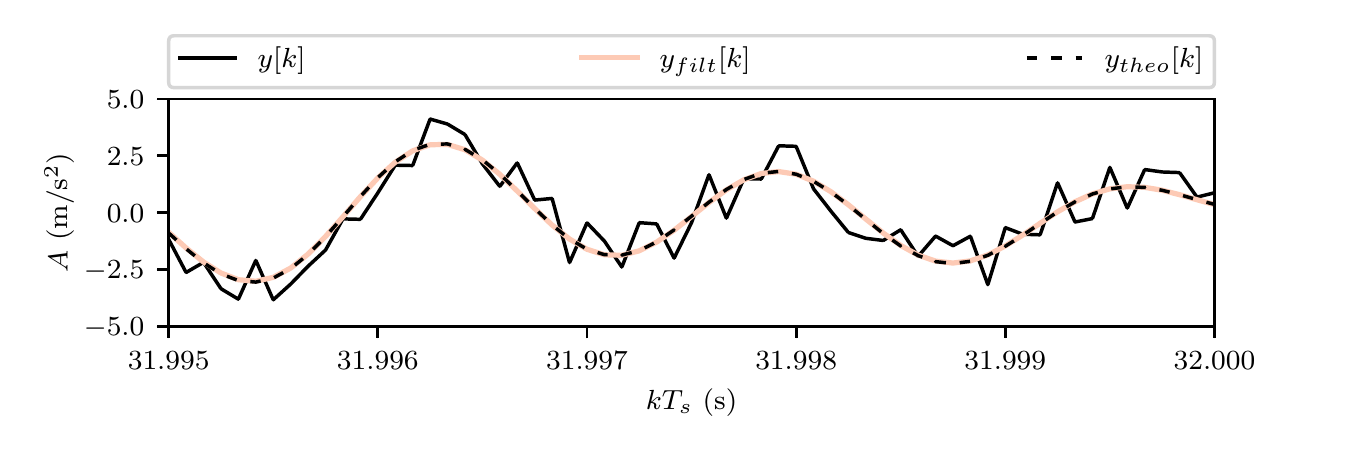}
    \caption{Zoom in view illustrating the noisy signal $y$ and the very good agreement between the filtered signal $y_{filt}$ and the theoretical signal $y_{theo}$.} \label{fig:1c}
  \end{subfigure}
\caption{Time series of the noisy signal $y[k]$ and of the noise free signal $y_{filt}[k]$ which is composed of the sum of $X_1[k]$, $X_2[k]$ and $X_3[k]$ which were extracted with the Vold-Kalman filter.} \label{fig1}
\end{figure}

Figure~\ref{fig:1a} illustrates the time series of the noise corrupted signal $y$, the noise-free signal $y_{filt}$ which is composed of the sum of the shaft signals $X_n$ extracted with the VKF.
Fig.~\ref{fig:1b} further shows the very good agreement between the resulting separated time series of the non-stationary components $X_n$ and the theoretical time series. The noise free signal $y_{filt}$ illustrated in Fig.~\ref{fig:1c} is then reconstructed from these extracted signals. One can observe that the extracted orders $X_n$ very closely match the theoretical ones. The VKF enables an order extraction in the time domain, but at a higher computational complexity than the Fourier transform. Its main advantage is that it allows the accurate separation of periodic signals with time varying amplitude, frequency and phase components from the remaining noise.

\section{Case study on real-world ABA measurements}
\label{sec:CSonABA}
\subsection{Data description}
\begin{figure}
\centering
\includegraphics[width=0.6\textwidth, angle =0,trim={0 0.0cm 0 2.2cm},clip]{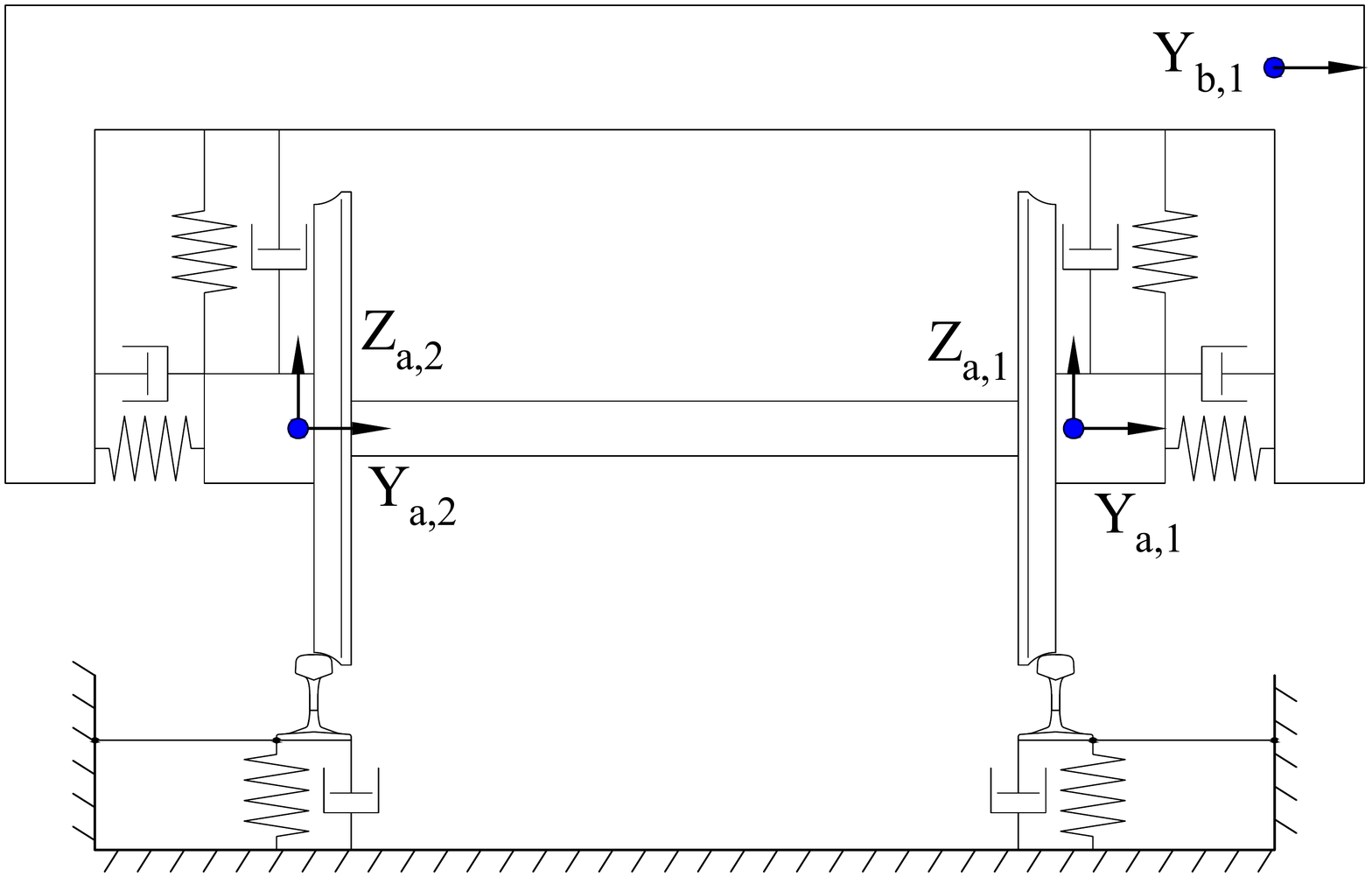}
\caption{Schematic 2D primary suspension representation of wheelset, bogie and track with corresponding measured degrees of freedom.} \label{fig00}
\end{figure}

The Swiss Federal Railways (SBB) have recently introduced the gDFZ, a specialized measurement vehicle equipped with many measurement systems, including ABA and tensiometric wheelsets (TWS). Tensiometric wheelsets are special wheelsets to measure wheel-rail forces \cite{schwabe_berg_2007}. The ABA and TWS are mounted to the first and the last axle (named axle~1 and axle~4 respectively) \cite{HoelzlReview2021}. 
Fig.~\ref{fig00} shows a simplified 2D representation of the vehicle axle connected to track and bogie. The sensors are positioned on the axle and the bogie. 
The sensor naming on railway vehicles usually follows the following convention: 
\begin{equation}\label{eq:sensornamingconvention}
Sensor \,Direction[Y,Z]_{Vehicle \,Axle,Vehicle \,Side}
\end{equation}
where the $Sensor \,Direction$ can be lateral ($Y$), vertical ($Z / Q$), where the $Vehicle \,Axle$ is the number of the axle (i.e. front axle $1$, trailing axle $4$) and the $Vehicle \,Side$ can be left ($2$) or right ($1$).
Data from several measurements covering more than 100~km of track on the network of the SBB are analyzed in this paper. ABA and rail--wheel force measurements from the \textit{gDFZ} vehicle are available since spring 2019.
In addition, other external sources of data are used here; a subsidence measurement was carried out on the same track section in 2016 by a specialized subsidence measurement vehicle (EMW, InfraMT \& SBB). Furthermore, the SBB fixed asset database (DfA) contains information on all track infrastructure components and their maintenance history, which allows for clustering of vehicle response characteristics by track type.
\newlength{\oldintextsep}
\setlength{\oldintextsep}{\intextsep}
\setlength\intextsep{2pt}

\subsection{Axle excitation frequencies}
The axle vibration response is composed of several periodic processes driven by the Out-Of-Roundness (OOR) of the wheel and the periodicity of sleepers. 
Such periodic processes are separated from non-deterministic components of the measured ABA signal using the VKF procedure outlined in Section \ref{sec:VKF}.
For structural health monitoring applications, separate classes of indicators can be derived from the deterministic component extracted by the VKF. Firstly, the angular velocity of the wheel $f_w[\ell=1]$ which is used in Eq.~\ref{eq:3} to extract indicators of the wheel condition (OOR and flat spots) is formulated as:
\begin{equation}\label{eq:25}
f_w[\ell] =\frac{v(t)\ell}{\pi\cdot d_{w}}
\end{equation}
where the vehicle speed $v(t)$ and the diameter of the wheel $d_{w}=0.92~m$ are known. Each order $\ell$ of wheel OOR causes an instantaneous frequency of $f_w[\ell=1,2,...n]$. 
Secondly, the track stiffness is related to the instantaneous frequency caused by the sleeper passage $f_s$, which is formulated as:
\begin{equation}\label{eq:26} 
f_s = f_{w}[1] \cdot \frac{\pi\cdot d_{w}}{d_{s}}
\end{equation}
where $d_{s}=0.6~m$ is the sleeper spacing.

Finally, the remaining signal components, which comprise non-deterministic processes and measurement noise that may be used to characterize all non-periodic effects.

\subsection{Order extraction with Vold-Kalman filter}
\begin{figure}
\centering
    \begin{minipage}[c]{0.849\linewidth}
      \begin{subfigure}{\textwidth}
        \includegraphics[trim={0.2cm 4.8cm 2cm 0.2cm},clip]{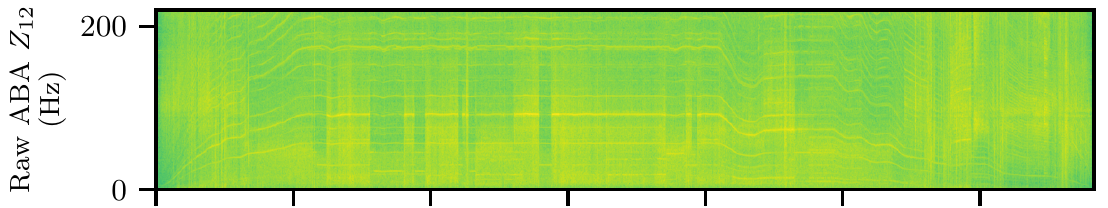}
        \centering
        \caption{} \label{fig:3a}
      \end{subfigure}
      \begin{subfigure}{\textwidth}
        \includegraphics[trim={0.2cm 2.5cm 2cm 2.3cm},clip]{figs/Spectrogram_Z12.pdf}
        \centering
        \caption{} \label{fig:3b}
      \end{subfigure}
      \begin{subfigure}{\textwidth}
        \includegraphics[trim={0.2cm 0.2cm 2cm 4.7cm},clip]{figs/Spectrogram_Z12.pdf}
        \centering
        \caption{} \label{fig:3c}
      \end{subfigure}
    \end{minipage} 
    \begin{minipage}[c]{0.14\textwidth}
      \begin{subfigure}{\textwidth}
        \includegraphics[trim={11.46cm 0.2cm 0.2cm 0.2cm},clip]{figs/Spectrogram_Z12.pdf}
      \end{subfigure}
    \end{minipage} 
\caption{a) Spectrogram of the raw ABA signal $Z_{12}$. b) Spectrogram illustrating the extracted VKF components for sensor channel $Z_{12}$. c) Non-stationary vehicle speed} \label{fig3}
\end{figure}

Harmonic signals are extracted from the ABA signals to account for the sleeper passage frequency and 11 orders of wheel out of roundness as defined in Tab.~\ref{tab:frequencies}. Fig.~\ref{fig3} illustrates the spectrogram of the original noisy signal and of the extracted harmonics as a function of the vehicle speed. The VKF separates periodic (harmonic) from non-periodic excitation components. The excitation components that correspond to the wheel geometry can be considered as constant over time and differences in the extracted envelope function between sensor locations may be used to estimate the wheel out of roundness.
The response to the excitation stemming from the passage of the sleepers that are spaced usually at a distance of $60\,[cm]$, is non-stationary, since the track properties (stiffness, inertia, damping) are varying.

\begin{table}[h]
\caption{Summary of VKF frequencies 
for wheel OOR and sleeper passage signal extraction}\label{tab:frequencies}%
\centering
\small
\begin{tabular}{lcc}
\toprule
VKF parameters &   Wheel OOR &  Sleeper passage \\
\midrule
$f$ (Hz) &    $f_w(l=1,...,11)$ (Eq.~\ref{eq:25}) & $f_s$ (Eq.~\ref{eq:26})\\
\bottomrule
\end{tabular}
\end{table}

\subsection{Wheel Out-Of-Roundness}
\begin{figure}
\centering
  \begin{subfigure}{0.39\textwidth}
    \includegraphics[trim={0.0cm 0.0cm 3.3in 0.0cm},clip]{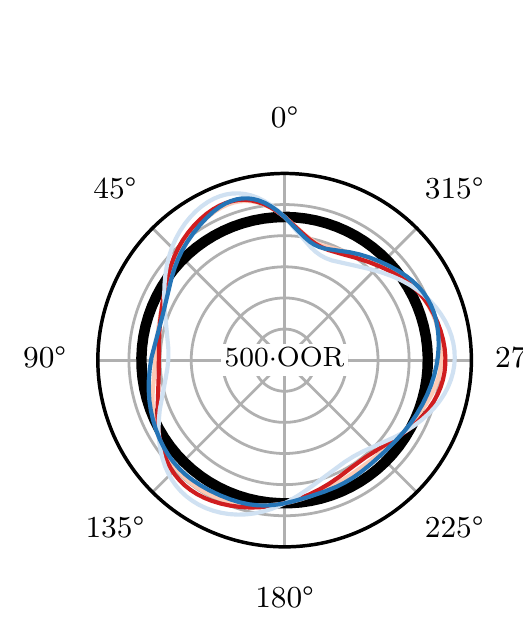}
    \caption{Polar plot of the wheel OOR} \label{fig:31a}
  \end{subfigure}
  \begin{subfigure}{0.59\textwidth}
    \includegraphics[trim={2.35in 0.0cm 0 0.0cm},clip]{figs/PolarPlot_OOR.pdf}
    \caption{Wheel OOR for the position along the circumference} \label{fig:31b}
  \end{subfigure}
\caption{Wheel OOR profile from the sum of the first 11 OOR orders.} \label{fig31}
\end{figure}

The wheel OOR causes a harmonic excitation, whose components are apparent in the acceleration spectrogram of sensor \textit{$Z_{12}$} in Fig.~\ref{fig3}. 
The profile of the wheel can be reconstructed by summing up the components of the 11 extracted VKF--orders that correspond to the wheel's OOR harmonic frequencies:
\begin{equation}\label{eq:27}
r(x) = 0.5\;d_w + \sum_{k=1}^{11}A_{k}(x)
\end{equation}
where the radius of the wheel $r$ is the sum of the mean wheel radius $r_w=0.5\cdot d_w=0.46$~m and the amplitude of the envelope function $A_{k}$ for each wheel OOR order $k$ along a position $x$ of the wheel circumference.
Fig.~\ref{fig31} shows a polar plot with the wheel OOR for the four wheels that are equipped with vertical acceleration sensors. The maximum OOR amplitude lies below 300~$\mu$m, which is equivalent to the OOR observed in the field tests by Nielsen and Johansson \cite{Nielsen2000}.
The left $Z_{42}$ and right  $Z_{41}$ wheels  on axle 4 have a similar OOR to the left $Z_{12}$ and right $Z_{11}$ wheels on axle 1.
While an exact profile is not available for the wheelsets, it is known that the wheelsets are in good condition at the time of measurement. The good quality of the condition can be confirmed by the low amplitude of the VKF-based profile reconstruction.

\subsection{Track stiffness}
\begin{figure}
\centering
      \begin{subfigure}{0.47\textwidth}
        \includegraphics[trim={0.15cm 0.0cm 2.82in 0.0cm},clip]{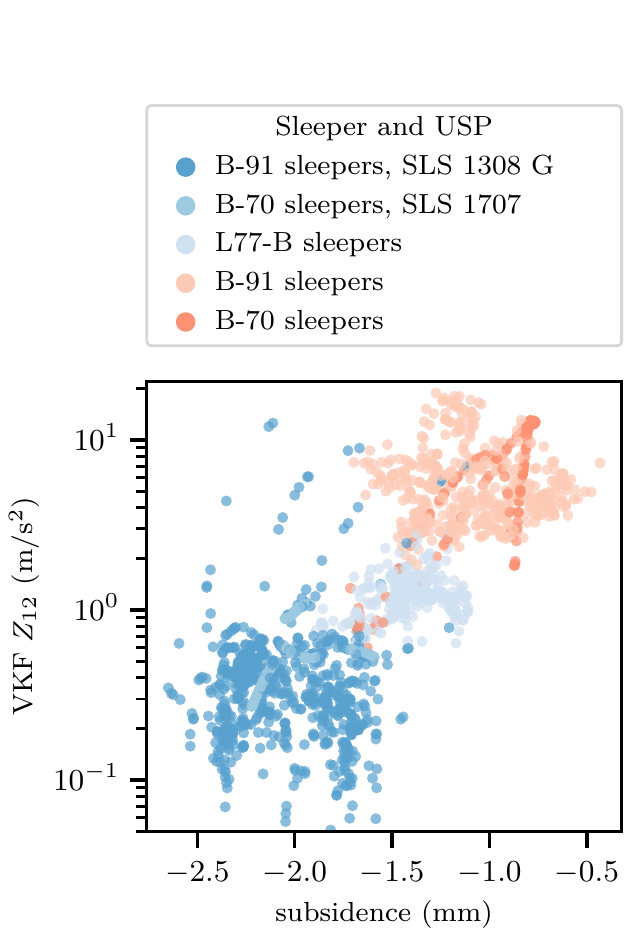}
        \caption{Scatter of VKF acceleration and subsidence} \label{fig:4a}
      \end{subfigure}\hfill
    \begin{minipage}[c]{0.53\linewidth}
      \begin{subfigure}{\textwidth}
        \includegraphics[trim={2.56in 1.96in 0 0.0cm},clip]{figs/ScatterHistogram_StiffnessAcceleration.pdf}
        \centering
        \caption{Subsidence along track position} \label{fig:4b}
      \end{subfigure}
      \begin{subfigure}{\textwidth}
        \includegraphics[trim={2.56in 0.0cm 0 1.95in},clip]{figs/ScatterHistogram_StiffnessAcceleration.pdf}
        \centering
        \caption{VKF $Z_{12}$ Acceleration along track position} \label{fig:4c}
      \end{subfigure}
    \end{minipage} 
\caption{The subsidence (inversely proportional to the track stiffness) and $Z_{12}$ acceleration amplitude are well correlated to each other. The VKF acceleration is higher on stiff sections (with low subsidence) which is primarily caused byvarying superstructure types.} \label{fig4}
\end{figure}

The amplitude of the sleeper passage component of the axle vibration varies in function of the vehicle speed and track stiffness. The vehicle-track system has several resonant modes, such as the pinned-pinned rail vibration at around 1000~Hz and the in-phase vibration of rail and sleepers at around 100~Hz \cite{Blanco2019}. As the vehicle speed varies, the parametric excitation due to the vehicle or the track may overlap with the frequency of the vibration modes, resulting in increased accelerations.
Fig.~\ref{fig4} shows acceleration samples collected during measurement rides performed at up to  200~km/h and the corresponding subsidence measured by the EMW. The three main clusters on the track section correspond to concrete sleepers with padding (B-91 sleepers with Sylomer SLS 1308G under sleeper pads), slab tracks (L77-B sleeper) and concrete sleepers with under-sleeper pads (B-91 sleepers without under sleeper pads). Since all four wheels equipped with ABA sensors show an equivalent response, the response for only $Z_{12}$ is illustrated here. Fig.~\ref{fig4} illustrates that when the subsidence approaches -0.5~mm, the acceleration amplitude of the in-phase sleeper passage mode $Z_{12}$ increases. Indeed, when the track stiffness is high (or equivalently when the subsidence is low), the contribution to the energy absorption of the passing vehicle mostly relies on the dynamics of the rail and the wheel, resulting in higher accelerations. The track subsidence $u$ is in fact proportional to the logarithm of the acceleration amplitude $Z_{12}$:
\begin{equation}
    u = a\cdot log(Z_{12})+b
\end{equation}
where the slope $a$ and the intercept $b$ are summarized in Tab.~\ref{tab:01}. The standard deviation of the residuals, also called the Root Mean Square Error (RMSE), is under 0.35~mm for all channels. 


   

The correlation between subsidence and VKF is not perfect because the EMW subsidence measurement was performed in 2015, while the earliest acceleration measurements are available from 2019. The difference can partly be explained by the different measurement processes; the VKF amplitude reflects the dynamic stiffness occurring at the regular line speed, while the subsidence measurements reflects the static stiffness that occurs at low speeds. In contrast, for the more flexible superstructures, the sleeper passage acceleration amplitude is lower as the load is more evenly distributed over multiple sleepers. As a result, the rail--wheel and sleeper-ballast contact forces decreases, which results in a longer track life \cite{Tzanakakis2013}. 

\begin{table}[h]
\caption{Subsidence prediction model}\label{tab:01}%
\centering
\small
\begin{tabular}{lrrrr}
\toprule
{Regressor} &   Wheel 11 &   Wheel 12 &   Wheel 41 &   Wheel 42 \\
\midrule
Incercept  &     -1.624 &     -1.617 &     -1.701 &     -1.760 \\
Slope      &      0.304 &      0.285 &      0.356 &      0.343 \\
RMSE       &      0.324 &      0.311 &      0.350 &      0.333 \\
Samples    &      18605 &      18605 &      1860  &      18605 \\
\bottomrule
\end{tabular}
\end{table}
\raggedbottom

\subsection{Relation between acceleration, forces and damage accumulation}\label{sec:relforceacc}
\begin{figure}
\centering
    \begin{minipage}[c]{0.455\textwidth}
      \begin{subfigure}{\textwidth}
        \includegraphics[trim={0.2cm 0.0cm 2.88in 0.0cm},clip]{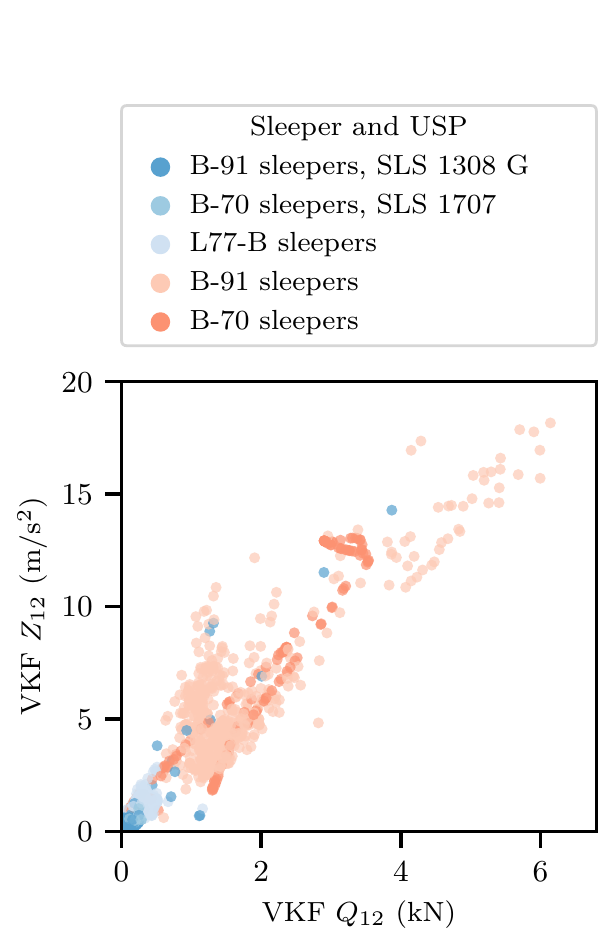}
        \caption{Scatter of forces and acceleration} \label{fig:5a}
      \end{subfigure}
    \end{minipage} 
    \begin{minipage}[c]{0.53\linewidth}
      \begin{subfigure}{\textwidth}
        \includegraphics[trim={2.55in 1.96in 0 0.0cm},clip]{figs/ScatterHistogram_ForceAcceleration.pdf}
        \centering
        \caption{Forces along track position} \label{fig:5b}
      \end{subfigure}
      \begin{subfigure}{\textwidth}
        \includegraphics[trim={2.55in 0.0cm 0 1.95in},clip]{figs/ScatterHistogram_ForceAcceleration.pdf}
        \centering
        \caption{Accelerations along track position} \label{fig:5c}
      \end{subfigure}
    \end{minipage} 
\caption{The VKF sleeper passage $Q_{12}$ force and $Z_{12}$ acceleration amplitude are well correlated to each other. The VKF force and acceleration amplitude show similar responses for varying superstructure types.} \label{fig5}
\end{figure}

The previous section showed that the amplitude of vibration of the unsprung wheel mass is determined by the track stiffness. In this section, the same VKF filtering process is applied on the tensiometric force measurements to extract the sleeper passage force amplitude. The direct relation between rail-wheel contact forces and ABA is important, as the vertical rail-wheel contact forces are a more direct indication for wear and fatigue on the infrastructure. Fig.~\ref{fig5} illustrates the direct relation between the amplitude of rail--wheel force and ABA amplitudes. The resulting measured forces $Q_{12}$ are proportional to the observed acceleration amplitudes $Z_{12}$. 
The unsprung wheel mass of around 300~kg is determined from the slope of the regression between the VKF sleeper passage amplitude of the force and the acceleration measurements. 

This direct relation between forces and accelerations is useful in quantifying track deterioration, since the degradation of the alignment of ballasted tracks is a time-dependent process, that originates from the dynamic excitation of vehicles. Repeated loading cycles, during which energy is dissipated via mechanical processes (damping of pads, friction of ballast grains,etc) cause material degradation. The consequence of the degradation of the ballast and the substructure is a reduced bearing capacity, leading to geometric deviations and settlement of the track structure \cite{Steenbergen2016}\cite{Steenbergen2013}. A settlement equation that relates the dynamic forces to the incremental ballast settlement is commonly used to model the degradation process \cite{Saussine2006}. The main parameters of such settlements models are the dynamic response forces on the ballast and the number of load cycles. 

\begin{figure}
\centering
  \begin{subfigure}{0.62\textwidth}
    \includegraphics[trim={0.0cm 0.1in 2.1in 0.0cm},clip]{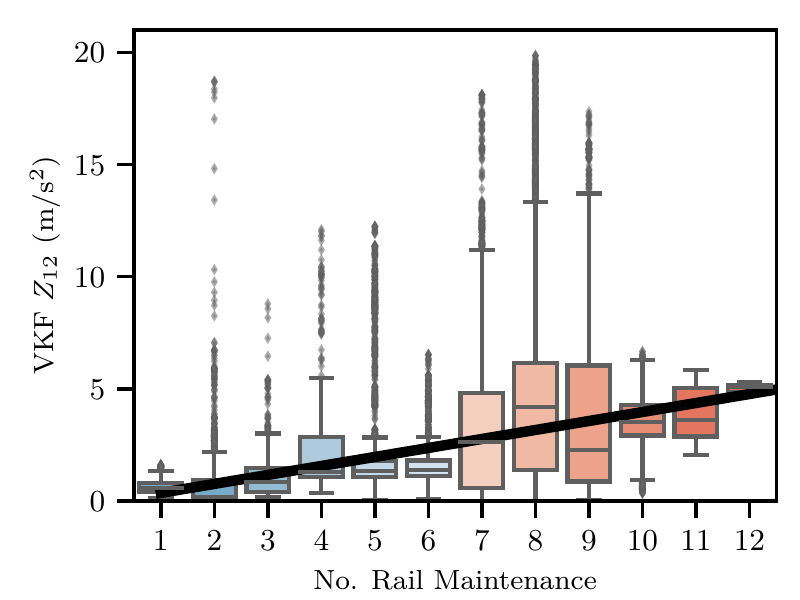}
    \caption{The VKF sleeper passage acceleration amplitude correlates to the number of rail maintenance actions.} \label{fig:6a}
  \end{subfigure}
  \begin{subfigure}{0.37\textwidth}
    \includegraphics[trim={3.2in 0.1in 0.1in 0.0cm},clip]{figs/Boxplot_Acceleration_MaintenanceRailBallast.pdf}
    \caption{VKF $Z_{12}$ correlates to the number of tamping actions.} \label{fig:6b}
  \end{subfigure}
\caption{Boxplots and regression line with amplitude of acceleration $Z_{12}$ in function of the number of rail maintenance actions and tamping since 2006 for a sampling interval of 25~cm on 100~km track sections.} \label{fig6}
\end{figure}

Hence, the damage accumulation to the infrastructure components is closely related to the load; larger load variations cause a higher rate of damage accumulation. 
In order to verify the relationship between the extracted indicators and the prompting for suitable maintenance actions, the track maintenance actions recorded in the DfA, since 2006, are extracted for the analyzed track segments for a total length of more than 100~km. 
The number of track maintenance actions that are related to the ballast and the rail are subsequently derived along the track at a spatial interval of 25~cm. Fig.~\ref{fig6} shows that the number of maintenance actions (tamping or grinding) correlates to the VKF-derived sleeper passage amplitude. The large variance in Fig.~\ref{fig6} is explained by two main factors. Firstly, maintenance is usually performed on a large scale and therefore these actions overlap between degraded and healthy sections. Secondly, preventive maintenance actions are not considered and may also influence the evolution of the overall track condition.

Fig.~\ref{fig5} demonstrates that the use of under sleeper pads has a beneficial effect by significantly lowering rail--wheel force amplitudes. This is confirmed by the work of Dahlberg \cite{Dahlberg2010}, who suggests that the wheel/rail contact forces are considerably reduced by decreasing the local stiffness variation via use of grouting or under-sleeper pads.

\section{Conclusion}
\label{sec:Concl}
The efficient and continuous monitoring of vehicle--track dynamics is essential for maximizing the lifespan of the railway infrastructure, while minimizing the required maintenance actions.
The Vold-Kalman Filter, a time-domain filtering technique, is here suggested as a fitting approach to extract condition indicators that characterize the condition of railway tracks and vehicle wheels. The multi-order extraction scheme separates the deterministic (periodic) components from non-deterministic contributions to the measured axle response.

The VKF amplitude components corresponding to the wheels are used to reconstruct the wheel profile, which in turn provides information on the condition of the wheel and the potential existence of flaws, such as wheel OOR. The VKF component related to the sleeper passage, on the other hand, contains information that is indicative of changes in the track stiffness.
It is shown that the VKF sleeper passage acceleration amplitude is proportional to the rail--wheel forces by a factor corresponding to the unsprung wheel mass. The VKF-derived indicator, thus, delivers an indirect, yet reliable, means to assess the underlying track stiffness, which serves as a proxy of the track condition. The VKF-based indicator can be exploited to improve the planning of optimal maintenance measures, since, railway track sections with higher stiffness were demonstrated in this and prior works to lead to increased forces at the rail--wheel interface, resulting in damage accumulation on the rails and ballast. Future work will explore the coupling of the VKF-derived stiffness indicator with existing substructure condition indicators, such as fractal values \cite{Landgraf_Enzi_Smartdata_proactive_Railway_Asset_Management}, which could further support he characterization of the specific source of increased stiffness. 

\section{Acknowledgements}
This work is supported by the Swiss Federal Railways (SBB) as part of the ETH Mobility Initiative program, under project OMISM - On board Monitoring for Integrated Systems Understanding \& Management Improvement in Railways. 
This research project is carried out in close cooperation with the Metrology (MUD) and Strategic Asset Management of the Track departments (SAFB) of the SBB. We would like to thank, in particular, our partners, Oliver Schwery and Stanislaw Banaszak, for their continuous support in project coordination.

\bibliographystyle{elsarticle-num} 
\bibliography{my}

\end{document}